\documentclass[fleqn,usenatbib]{mnras}
\usepackage[T1]{fontenc}
\usepackage{amssymb,epsfig}
\usepackage[]{empheq}
\usepackage{natbib,ifthen}
\usepackage{mathrsfs}
\usepackage{color}
\usepackage{bbold}
\usepackage{csvsimple}
\usepackage{hyperref}       
\usepackage{url}            
\usepackage{booktabs}       
\usepackage{amsfonts}       
\usepackage{amsmath}
\usepackage{graphicx}
\usepackage{wrapfig}
\usepackage{braket}
\usepackage{physics}
\usepackage{algorithm2e}

\usepackage{xcolor}

\newcommand{\x}{\boldsymbol{x}}

\newcommand{\s}{\boldsymbol{s}}

\title[Periodicity significance testing with null-signal templates]{
Periodicity significance testing with null-signal templates: reassessment of PTF's SMBH binary candidates
}

\author[J. Robnik, et al.]{
Jakob Robnik,$^{1}$\thanks{E-mail: jakob\_robnik@berkeley.edu}
Adrian E.~Bayer,$^{2,3}$
Maria Charisi,$^{4,5}$
Zolt\'an Haiman,$^{6,7}$
Allison Lin,$^{6}$
Uro\v{s} Seljak$^{1,8}$
\\
$^{1}$Department of Physics,
University of California, Berkeley, CA 94720, USA\\
$^{2}$Department of Astrophysical Sciences, Peyton Hall, Princeton University, Princeton, NJ 08544, USA\\
$^{3}$Center for Computational Astrophysics, Flatiron Institute, 162 5th Avenue, New York NY 10010, USA\\
$^4$Department of Physics and Astronomy, Washington State University, Pullman, WA 99163, USA\\
$^5$Institute of Astrophysics, FORTH, GR-71110, Heraklion, Greece\\
$^{6}$Department of Astronomy, Columbia University, New York, NY 10027, USA\\
$^{7}$Department of Physics, Columbia University, New York, NY 10027, USA\\
$^{8}$Lawrence Berkeley National Laboratory, 1 Cyclotron Road, Berkeley, CA 93720, USA
}

\pubyear{2020}

\begin{document}
\label{firstpage}
\pagerange{\pageref{firstpage}--\pageref{lastpage}}
\maketitle

\begin{abstract}
    Periodograms are widely employed for identifying periodicity in time series data, yet they often struggle to accurately quantify the statistical significance of detected periodic signals when the data complexity precludes reliable simulations (e.g., in the presence of the correlated and/or non-Gaussian noise). We develop a data-driven approach to address this challenge by introducing a null-signal template (NST). The NST is created by carefully randomizing the period of each cycle in the periodogram template, rendering it non-periodic. It has the same frequentist properties as a periodic signal template regardless of the noise probability distribution, and we show with simulations that the distribution of false positives is the same as with the original periodic template, regardless of the underlying data. 
    Thus, performing a periodicity search with the NST acts as an effective simulation of the null (no-signal) hypothesis, without having to simulate the noise properties of the data.
    We apply the NST method to the supermassive black hole binaries (SMBHB) search in the Palomar Transient Factory (PTF), where Charisi et al. had previously proposed 33 high signal to (white) noise candidates utilizing simulations to quantify their significance. Our approach reveals that these simulations do not capture the complexity of the real data. There are no statistically significant periodic signal detections above the non-periodic background. To improve the search sensitivity we 
    introduce a Gaussian quadrature based algorithm for the Bayes Factor with correlated noise as a test statistic, in contrast to the standard signal to white noise. We show the method is accurate and much faster than MCMC, which enables the computation of the Bayes Factor for all the quasars in the PTF sample. We show with simulations that this improves sensitivity to true signals by more than an order of magnitude. However, using the Bayes Factor approach also results in no statistically significant detections in the PTF data. 
\end{abstract}

\begin{keywords}
quasars: supermassive black holes, 
methods: statistical,
software: data analysis
\end{keywords}

\section{Introduction}

Detecting periodicity in time series data $X$ is an important task in many scientific applications, ranging from astronomy to economics. 
In the frequentist approach to hypothesis testing, one constructs a test statistic $q(X)$, which is designed to be large only when a periodic signal is present in the data.
For example, for irregularly sampled time series, the Lomb-Scargle periodogram score is a common test statistic \citep{lomb_least-squares_1976, scargle_studies_1982, vanderplas_understanding_2018}. For white noise containing a sinusoidal signal, the Lomb-Scargle score equals the likelihood ratio between the signal and no-signal hypotheses. However, for more complex noise and signals the correspondence to the likelihood ratio no longer holds, rendering it a sub-optimal test statistic.

A test statistic is calibrated by its distribution under the null hypothesis, $\mathcal{H}_0$, i.e. the hypothesis that there is no signal. 
A signal is detected if the p-value = $P(q > q(X) \vert \mathcal{H}_0)$ is sufficiently small.
Typically, multiple simulations of the data under the null hypothesis are generated and the p-value is estimated as the fraction of simulations for which $q > q(X)$. 
This approach is robust to the assumptions about the alternative hypothesis, i.e. hypothesis that the signal is present, and the choice of the test statistic: sub-optimal choices result in a lower probability of detecting the signal, but one still can be confident about the false positive probability.
However, the approach is not robust to the assumptions about the null hypothesis.
Realistic datasets are often too complex to be accurately simulated, often leading to an underestimation of the p-value, which in turn leads to false detections.
The situation is even worse if one attempts to reduce the computational complexity and replace simulations with analytic estimates. For example, the analytic expressions provided in \cite{baluev_assessing_2008, baluev_impact_2013} are only valid for independent, identically distributed (i.i.d.) Gaussian noise. 
\cite{bayer_look-elsewhere_2020} and \cite{robnik_statistical_2022} proposed to estimate the p-value from the Bayes Factor, but this is also sensitive to assumptions of the underlying model, since the Bayes Factor depends on it.
 
Data-driven approaches to quantify the p-value help to avoid this issue.
For example, \cite{bayer_self-calibrating_2021} estimates the p-value of the heighest peak in the periodogram using the secondary peaks in the periodogram. 
An alternative is to directly use the data to create effective null ``simulations''. One such approach is bootstraping \citep{ivezic_lsst_2019}, where each permutation of the measurements is treated as a different realization of the null. However, this approach is only valid if the null is composed of independent and identically distributed data. 

In this work, we propose a novel data-driven approach for quantifying the false positive probability in periodicity searches based on the modification of the periodic search template, rather than the data. The null-signal template (NST) 
is constructed from the periodic template by stretching and compressing the individual cycles of the signal.
NST will assign low significance to the real (periodic) signal, but will have the same statistical properties as the original template on the null. Analysis with the null-signal template can therefore be used as an effective null simulation. This approach has been used in other fields. For example, gravitational waves trigger a coincident detection across multiple detectors, with a small delay, equal to the light crossing time between the detectors. Therefore, by fixing a large non-physical delay between the detectors, one can be sure to find only false positives \citep{ligo_scientific_collaboration_analysis_2004}. Their distribution is the same as the false positive distribution in the original search with coincident detection, and thus the search with the modified delay can be used as a substitute for the null hypothesis simulations. When applying the method to applications such as periodograms the challenge is to find 
NST that reproduces the same false positive rate
as the actual search, while also being 
sufficiently orthogonal to the presence of true signals such that the resulting 
false positive rate is only weakly affected by the possible existence of true signal in the data. 


We apply the method to the search for quasar periodicity, which can signify the presence of sub-parsec super-massive black hole binaries (SMBHBs). In recent years, systematic searches in large samples of quasar light curves from time-domain surveys have revealed around 150 candidates. For instance, \citet{graham_systematic_2015} identified 111 candidates in a sample of around 250,000 quasars from the Catalina Real-time Transient Survey (CRTS), while \citet{charisi_population_2016} identified 33 candidates among around 35,000 quasars from the Palomar Transient Factory (PTF). More recent searches in the Panoramic Survey Telescope and Rapid Response System (Pan-STARRS), the Dark Energy Survey (DES) and the Zwicky Transient Facility (ZTF) detected a few more candidate systems \citep{liu_supermassive_2019,chen_candidate_2020,chen_searching_2024}. For a detailed summary of the current status of quasar periodicity searches, we refer the reader to \citet{charisi_multimessenger_2022,dorazio_observational_2023} 
Here we apply our new method to reassess the periodicity significance of the 33 SMBHB candidates identified in PTF.

In addition to the NST method, which is agnostic to the underlying noise properties, we also introduce some improvements to the standard quasar periodicity detection methods, which typically model quasar variability as a damped random walk (DRW) \citep{macleod_modeling_2010, kelly_are_2009, kozlowski_quantifying_2010}.
Previous searches for quasar periodicity \citep{graham_systematic_2015, charisi_population_2016,chen_candidate_2020} used periodogram peaks as their test statistic and then assessed the p-values of the detected signals with DRW simulations. Given that the underlying assumption in the periodogram is uncorrelated white noise, this makes the choice of this statistic sub-optimal, as the noise is correlated. In addition, priors may also improve the search sensitivity, by focusing on the regions 
of parameter space with higher prior. 
The posterior odds,
often equivalent to the Bayes Factor, are known to be the optimal test statistic \citep{zhang_bayesian_2017, fowlie_neymanpearson_2023} in the sense that they maximize the true positive rate (TPR) at a fixed false positive rate (FPR). 
However, the Bayes Factor with correlated Gaussian noise has been computed only for a few particularly interesting candidates \citep{dorazio_relativistic_2015, zhu_toward_2020, vaughan_false_2016}. 
This is because the evidence integral is usually computed with sampling algorithms such as nested sampling \citep{zhu_toward_2020,witt_quasars_2022}, which is computationally intensive. 
Sometimes a Bayesian information criterion (BIC) is used as an easy to evaluate substitute for the Bayes Factor \citep{witt_quasars_2022, liu_did_2018}, but BIC does not take into account the prior information and accounts for the look elsewhere effect overly simplistically leading to a suboptimal analysis \citep{weakliem_critique_1999}.

Exploiting the low dimensionality of the problem, we instead compute the Bayes Factor with a Gaussian quadrature scheme \citep{robnik_statistical_2022}, achieving the same accuracy as sampling methods at about three orders of magnitude lower computational cost. Specifically, a single light curve analysis with our code takes around half a minute on a single CPU, while in \cite{witt_quasars_2022}, several CPU hours were used. The speed-up enables the use of the Bayes Factor to search for periodicity  in the entire quasar sample, which previously would have been prohibitive due to computational demands. This can be scaled to the even larger samples in future surveys, like  the Rubin Observatory \citep{ivezic_lsst_2019-1, xin_ultra-short-period_2021} or the Roman telescope \citep{haiman_massive_2023}, which will detect millions of quaasars. 

In Section \ref{sec: method} we present the methodology of the novel data-driven approach to use null-signal templates to create effective null simulations. In Section \ref{sec: synthetic} we validate the method on synthetic light curves. In Section \ref{sec: quasars} we then present the improvements to the SMBHB search in PTF data and apply the NST to calibrate the results. 
The code with tutorials is available at \url{https://github.com/JakobRobnik/periodax}.

\section{Method} \label{sec: method}

Let us consider a time-series $X$, consisting of measurements $\{ x_i \}_{i=1}^N$, with measurement errors $\{ \sigma_i \}_{i=1}^N$, taken at times $\{ t_i \}_{i=1}^N$, where $N$ is the total number of observations. Let $T$ be the time span of the data (i.e the baseline of the light curve). The measurements $x_i$ are composed of noise $n_i$ and signal $s(t_i)$, so that:
\begin{equation}
    x_i = s(t_i) + n_i,
\end{equation}
where the signal 
\begin{equation} \label{periodic}
    s(t) = A \, u \left( \frac{t}{P} - \phi \right)
\end{equation}
is periodic, i.e. $u(x + k) = u(x)$ for integer $k$. $P$ is the period, $\phi \in [0, 1)$ the phase, and $A$ the amplitude of the periodic signal. The standard periodogram corresponds to a simple sinusoidal signal, $u(x) = \sin \{ 2 \pi x \}$. Other examples include exoplanet transit searches, where $u(x)$ is a U-shaped transit \citep{robnik_matched_2021}, or more complex SMBHB signals, like sawtooth periodicity \citep{duffell_circumbinary_2020, westernacher-schneider_multiband_2022} or periodically repeating lensing flares \citep{dorazio_detecting_2020}.

We would like to know if a periodic signal is present in the data. The strategy is to design some test statistic $q(X)$. A popular test statistic for periodicity detection is the Lomb-Scargle periodogram score:
\begin{align} \label{eq: ls}
    q_{LS}(X \vert s) &= \max_{A, P, \phi} \, \sum_{i, j =1}^N x_i \Sigma^{-1}_{ij} x_j - \Delta_i \Sigma^{-1}_{ij} \Delta_j  \\ \nonumber
    &\Delta_i \equiv x_i - s(t_i \vert A, P, \phi) ,
\end{align}
where $\Sigma$ is the noise covariance matrix, which is diagonal in the standard white Lomb-Scargle periodogram, $\Sigma_{ij} = \delta_{ij} \sigma_i^2$, and only accounts for the measurement errors. 
Another common test statistic is the likelihood ratio 
\begin{equation} \label{eq: lr}
    q_{LR} = 2 \log \frac{\max_{\boldsymbol{z}_1}p(X \vert \mathcal{H}_1, \boldsymbol{z}_1)}{\max_{\boldsymbol{z}_0}p(X \vert \mathcal{H}_0, \boldsymbol{z}_0)},
\end{equation}
where $\boldsymbol{z}_i$ are the parameters of the $\mathcal{H}_i$ hypothesis. Lomb-Scargle score is a special case of the likelihood ratio, namely when the noise is Gaussian and the null has no parameters. While the likelihood ratio takes the ratio of the likelihoods at the optimal parameters, the
Bayes Factor takes the ratio of the evidences:
\begin{equation} \label{eq: bf}
    B = \frac{p(X \vert \mathcal{H}_1)}{p(X \vert \mathcal{H}_0)},
\end{equation}
which are the likelihoods averaged over the prior $p(X \vert \mathcal{H}_i) = \int p(X \vert \boldsymbol{z}_i) p(\boldsymbol{z}) d \boldsymbol{z}_i$. Finally, the posterior odds additionally weigh the Bayes Factor with the prior odds, i.e. the ratio $p(\mathcal{H}_1) / p(\mathcal{H}_0)$ that favors the datasets where the signal is more probable a priori:
\begin{equation} \label{eq: po}
    \frac{p(\mathcal{H}_1 \vert X)}{p(\mathcal{H}_0 \vert X)} = \frac{p(X \vert \mathcal{H}_1)}{p(X \vert \mathcal{H}_0)} \frac{p(\mathcal{H}_1)}{p(\mathcal{H}_0)}.
\end{equation}
%
Once a test statistic is chosen, it is evaluated on the data $q(X)$ and is compared to the distribution $P(q \vert \mathcal{H}_0)$ under the null hypothesis. If the observed value is very unlikely under the null, we can claim a discovery.

Here we propose to estimate the null distribution by modifying the test statistic $q(X) \xrightarrow[]{} \widetilde{q}(X)$, such that the statistical properties of both statistics are the same under the null:
\begin{equation} \label{condition 1}
    P(\widetilde{q} \vert \mathcal{H}_0) \approx P(q \vert \mathcal{H}_0) , \tag{C1}
\end{equation}
while the modifed statistic is only weakly triggered on the alternative
\begin{equation} \label{condition 2}
    P(\widetilde{q} \vert \mathcal{H}_1) \approx P(\widetilde{q} \vert \mathcal{H}_0) \ll P(q \vert \mathcal{H}_1). \tag{C2}
\end{equation}
A test statistic satisfying Conditions \eqref{condition 1} and \eqref{condition 2} can be used as an effective null simulation when applied to the real data, regardless of whether the data is actually from $\mathcal{H}_0$ or $\mathcal{H}_1$. This is a purely data driven approach, and does not require any 
simulations to quantify the false positive
rate. 

\begin{figure}
    \centering
    \includegraphics[scale = 0.21]{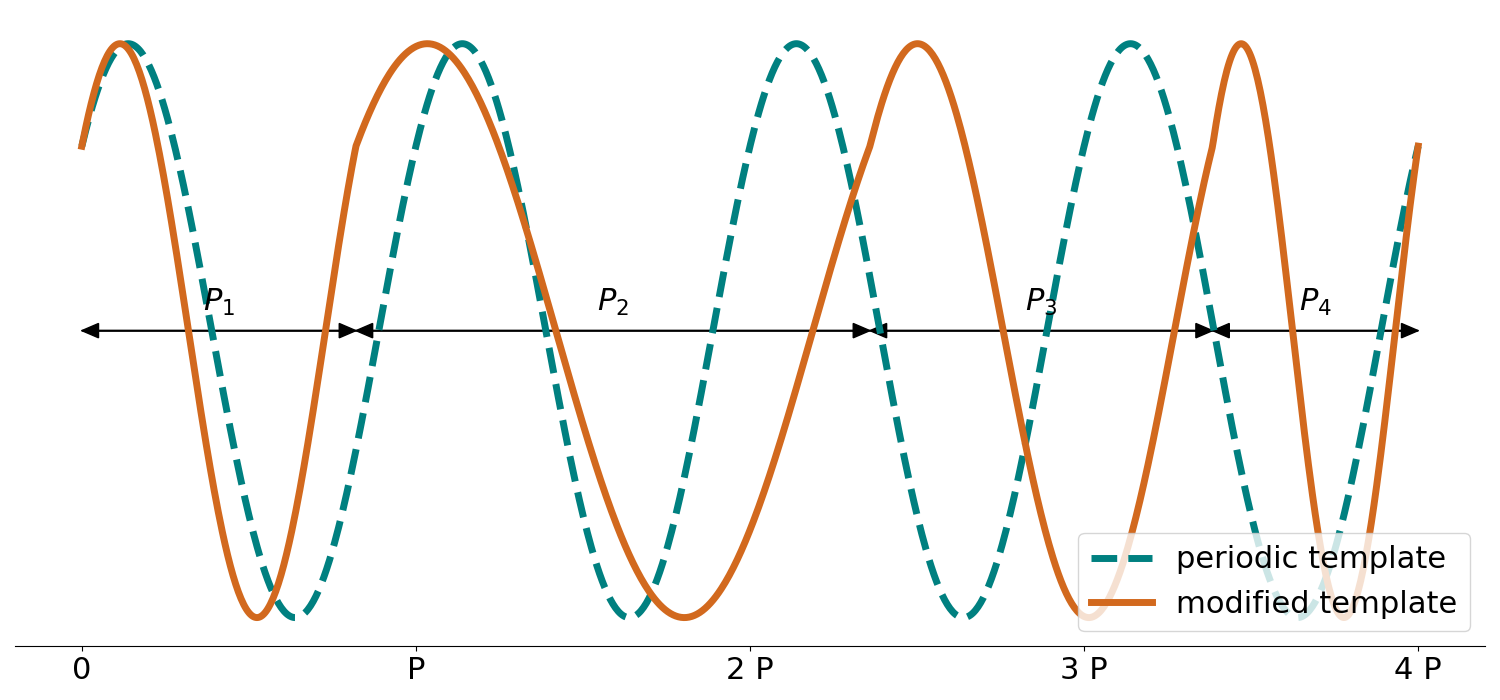}
    \caption{An example of a sinusoidal signal template (teal) and the null sinusoidal template (orange) from Eq. \eqref{eq: null signal template}. The individual cycles of the null-signal template are marked with arrows.}
    \label{fig: template}
\end{figure}

We construct $\widetilde{q}$ by perturbing the signal template of the alternative hypothesis $\mathcal{H}_1$, such that $\widetilde{q}(X) = q(X \vert \widetilde{s})$. For example, $\widetilde{q}_{LR}(X)$ is the ratio of the likelihoods, but the alternative hypothesis uses the perturbed template.
The null-signal template $\widetilde{s}$ is the same as the original template on short time scales, i.e. the period of the signal, but differs significantly on time scales comparable to the length of the light curve. 
The similarity of the templates on short time scales ensure that $q$ and $\widetilde{q}$ trigger similarly on the null signal, satisfying condition \eqref{condition 1}, while
long-scale differences ensure that the overlap between the null-signal template and the original template is small, hence satisfying condition \eqref{condition 2}. This method thus works best when the number of signal cycles is large, but we provide an additional prescription to apply it with as few as two signal cycles.

There are multiple ways of modifying the template; in this work, we randomize the length of its cycles. The original template is periodic, so the $n$-th cycle is the interval $[(n-1) P, \, n P ]$. The null-signal template's cycle $n$ will have length $P_n$, such that the $n$-th cycle now spans the interval $[\sum_{k = 1}^{n-1} P_k, \, \sum_{k = 1}^{n} P_k ]$. Inside each cycle, the null-signal template equals the stretched or contracted original template cycle.
More formally, the null-signal template is
\begin{equation} \label{eq: null signal template}
    \widetilde{s}(t) = A \, u \left( \frac{t - t_0(t)}{P_{i(t)}} - \phi \right),
\end{equation}
where $i(t) \equiv \mathrm{max} \{ j \vert \sum_{k =1}^j P_k < t\}$ and $t_0(t) \equiv \sum_{k = 1}^{i(t)}P_k$.
Note that this template is continuous if $u$ is continuous, but it is not differentiable. An example of the null-signal template is shown in Figure \ref{fig: template} in orange, while the original periodic template is shown in teal.

It remains to select the lengths $P_n$ of cycles. 
The original template Eq. \eqref{periodic} is a function of the signal period $P$, so an equivalent of the period must also be defined for the null-signal template Eq. \eqref{eq: null signal template}. Furthermore, the null-signal template must vary smoothly with the period. We achieve this by requiring that the null-signal template at period $P$ has the same number of possibly noninteger cycles as the original template with period $P$, which is $T/P$. This amounts to the following constraint
\begin{equation} \label{eq: number of cycles constraint}
    P_1 + P_2 + \ldots P_n + \lambda P_{n+1} = T = (n + \lambda) P .
\end{equation}
Here, the number of cycles is decomposed into integer and fractional parts: $T / P = \lfloor T/P \rfloor + \text{frac}(T/P) \equiv n + \lambda$. 
The constraint from Eq. \eqref{eq: number of cycles constraint} is satisfied if we pick
\begin{equation} \label{eq: normalization}
    P_n = \frac{T p_n}{\lambda p_{n+1} + \sum_{k = 1}^n p_k} ,
\end{equation}
where $\{ p_n \}$ can be any positive numbers. For example, one could use i.i.d. draws from the Gamma distribution which would result in periods being symmetrically Dirichlet distributed. We would like to avoid the tails of the Gamma distribution so we pick the uniform distribution, $p_n \sim \mathcal{U}(1/\sqrt{3}, \sqrt{3})$. 

\begin{figure}
    \centering
    \hspace*{-0.25cm}\includegraphics[scale = 0.35]{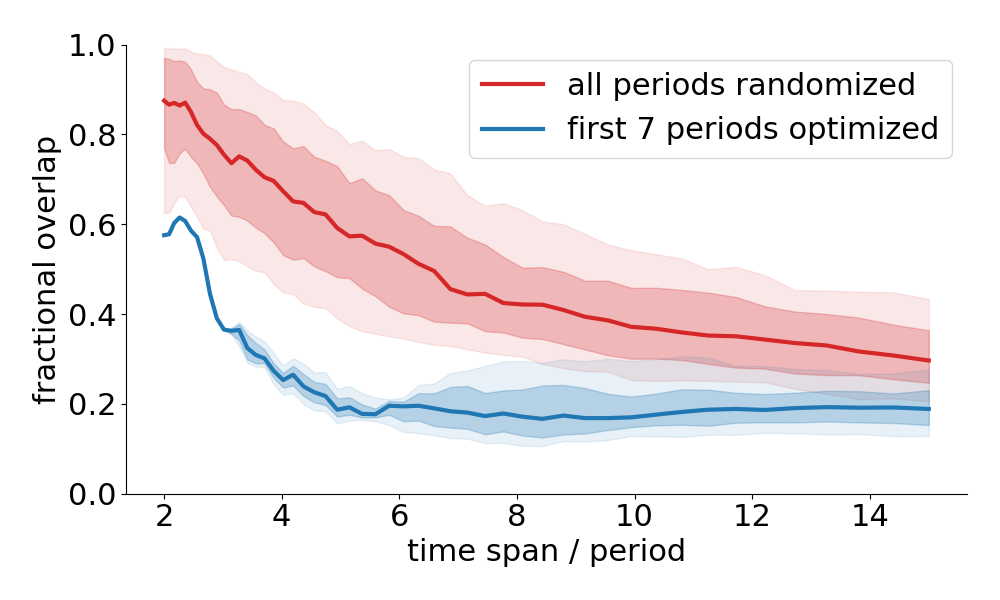}
    \caption{
    Lomb-Scargle periodogram score with the null-signal template (NST) relative to the score with the periodic template, i.e. fractional overlap $r(s_{\mathrm{NST}}, s_{\mathrm{periodic}})$. The data here are just the periodic template with no noise. Lower is better, because it ensures NST has small overlap with the periodic template and consequently obeys Condition \eqref{condition 2}.
    The red line corresponds to NST if all periods are drawn from the uniform distribution. The blue line uses the fixed first seven periods that minimize the overlap and random other periods, as described in Section \ref{sec: method}. Time sampling is 20 equally spaced points / cycle and the frequency grid for the NST is 1000 logarithmically spaced frequencies in the range $[P/10, 10 P]$.
    The solid lines are medians over two hundered NST period randomness realizations. The stronger and the weaker band correspond to 25-75 and 10-90 quantiles respectively.
    As can be seen the optimized periods (blue) correspond to the best possible case of the random periods (red) and reduce the overlap to $60 \%$ even at only two signal cycles.
    }
    \label{fig: overlap}
\end{figure}

Having random periods can be problematic if the number of cycles is very low, because all periods can become very similar to each other by pure chance. Should this happen, the null-signal template would no longer have small overlap with the periodic template and would violate Condition \eqref{condition 2}. To prevent this from happening we choose $p_n$ deterministically for low $n$.
We define the fractional overlap between two templates as $r(s_{\mathrm{test}}, s_{\mathrm{true}}) = q_{LS}(s_{\mathrm{true}} \vert s_{\mathrm{test}}) / q_{LS}(s_{\mathrm{true}} \vert s_{\mathrm{true}})$. This is the ratio of the Lomb-Scargle scores that the template $s_{\mathrm{test}}$ would assign to a signal which was actually generated by $s_{\mathrm{true}}$, relative to the score that we would get using the periodic template $s_{\mathrm{true}}$. 
Small values of the fractional overlap will ensure that Condition \eqref{condition 2} holds. We can achieve this by optimizing the values $p_n$.
We do this iteratively, starting with $p_1 = 1$ without loss of generality. In the $n$-th iteration, $p_1, p_2, \ldots p_n$ have already been fixed. We take the time span $T = (n+1) P$ and optimize $p_{n+1}$ such that the overlap $r(s_{\mathrm{true}}, s_{\mathrm{test}})$ is minimized. To make the optimized values more robust to the possible gaps in the data, we try $n$ different gap options, each covering one of signal cycle. The function that we then minimize is $r$, maximized over the gap choice. 
To compute the overlap we use a sinusoidal template with an almost continuous time sampling (20 equally spaced points / cycle) and a dense grid for the periodic template trial frequencies (1000 logarithmically spaced frequencies in the range $[P/10, 10 P]$). We optimize for the first 7 $p_n$ and normalize them to unit average to match the scale of $p_{n>7}$. These are then fixed throughout the paper.
Figure \ref{fig: overlap} shows that this procedure yields small overlaps, even when the time span of the data is not much larger than the period of the signal.

Note that this method only ensures that the fractional overlap between the periodic template and the NST is small, but not its absolute value. Therefore, the presence of a very large alternative hypothesis signal can compromise the first equality in condition \eqref{condition 2}. In this case, it may be better to perform iterative analysis, such as subtracting the best fit signal from the data and compute the modified test statistic on residuals, otherwise the p-value estimate would be conservative. Either way this case is easy to identify, and will not be an issue for the current work, since as we will show no periodic signal candidates surpass the NST baseline.

\section{Validation} \label{sec: synthetic}

\begin{figure}
    \centering
    \includegraphics[scale = 0.23]{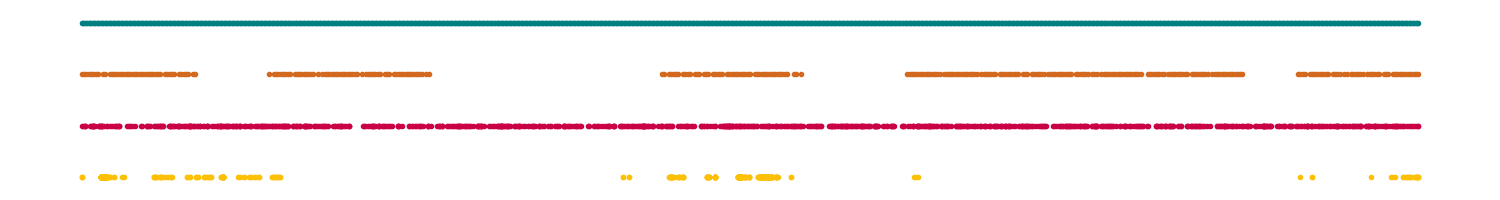}
    \caption{Time sampling in synthetic datasets from Section \ref{sec: synthetic}. Time is on the x-axis (but with different scales for different time samplings), the dots indicate times where measurements are taken. From the top to the bottom: equally spaced, equally spaced with gaps, randomly spaced. The bottom row is an example PTF time sampling for a quasar SDSS J085037.61+201337.1.}
    \label{fig:timesampling}
\end{figure}

\begin{figure*}
    \centering
    \includegraphics[scale=0.33]{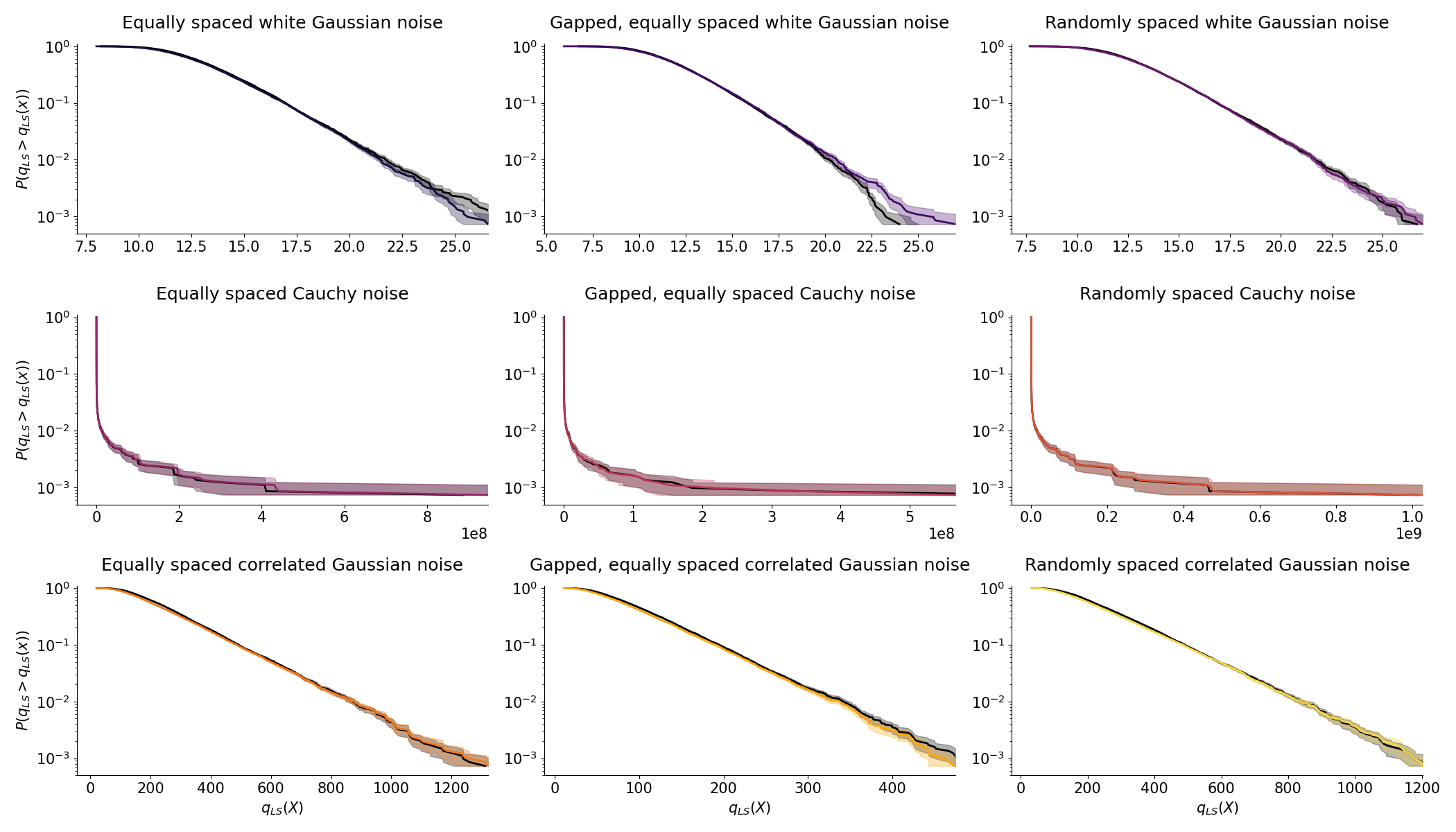}
    \caption{On noise-only simulations, the p-value with the null-signal template (colored) agrees perfectly with the periodic template (black). 
    A grid with all possibilities of time sampling and noise distribution from Section \ref{sec: synthetic} is shown. 
    This demonstrates that validity of Condition \eqref{condition 1} is not sensitive to the noise and time sampling properties. Note the large case to case differences in the signal scores.
    The shaded bands are statistical one $ \sigma$ uncertainties due to the finite number of simulations, obtained with bootstraping.}
    \label{fig:synthetic}
\end{figure*}

\begin{figure*}
    \centering
    \includegraphics[scale=0.33]{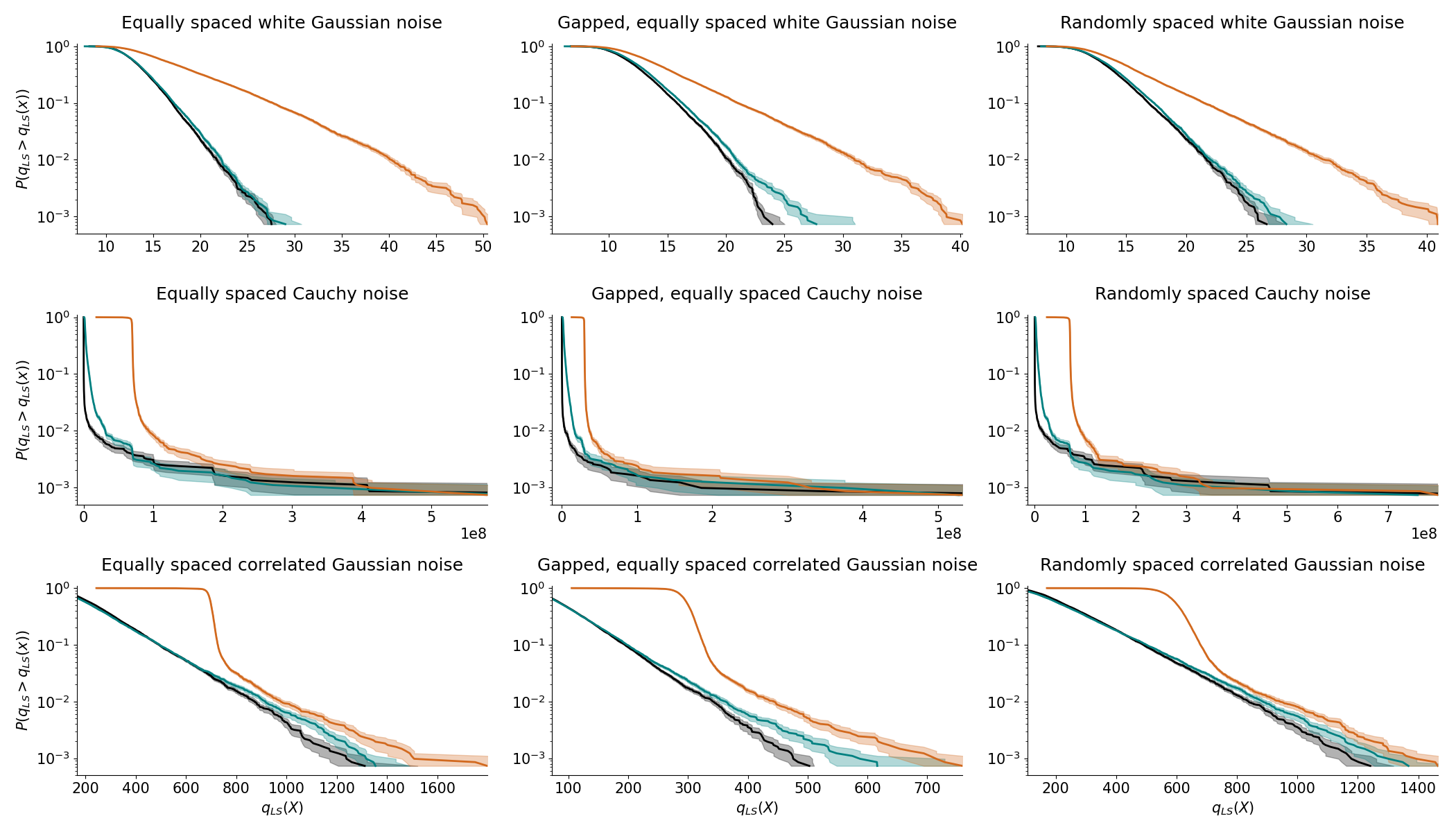}
    \caption{On noise with injected signal simulations, the score distribution with the null-signal template (blue) differs significantly from the periodic signal (orange) and matches almost perfectly the distribution with the periodic template on noise-only simulations (black). 
    This demonstrates that validity of Condition \eqref{condition 2} is not sensitive to the noise and time sampling properties.
    A grid with all possibilities of time sampling and noise distribution from Section \ref{sec: synthetic} is shown.}
    \label{fig:synthetic1}
\end{figure*}

We now test Conditions \eqref{condition 1} and \eqref{condition 2} in a variety of synthetic examples where exact null simulations can be performed and compared to the NST effective null simulations. 
The examples are chosen to provide some variety of noise properties and time samplings. We consider three different noise distributions:
\begin{itemize}
    \item i.i.d. standard Gaussian,
    \item i.i.d. Cauchy,
    \item correlated Gaussian with a DRW kernel, $\tau = T/20$, \\
    $\sigma = 1$ and measurement errors $\sigma_i = 0.2$. $T$ is the time span of the light curve.
\end{itemize}
with three different time samplings:
\begin{itemize}
    \item 1000 equally spaced observations, $t_k = k$,
    \item equally spaced with some observations missing. We 
    select four large regions where the data is missing and additionally independently remove each measurement 
    with 10 \% probability. This time sampling somewhat
    resembles Kepler space telescope sampling 
    \citep{jon_m_jenkins_kepler_2017}
    \item 1000 uniformly random sampled $t_k \sim \mathcal{U}(0, 1)$.
\end{itemize}
Time samplings are shown in Figure \ref{fig:timesampling}.
For each of the 9 combinations of time samplings and noise, we do $8192$ simulations. For the test statistic we use the floating mean periodogram score from Appendix \ref{sec: mf}, maximized over the trial frequencies $\{ k/T \}_{k =1}^{N/2}$. 
The resulting cumulative distribution with the sinusoidal and the null template are shown in Figures \ref{fig:synthetic} and \ref{fig:synthetic1}. 
In Figure \ref{fig:synthetic} we perform noise-only simulations and show a perfect match between the two templates. Thus Condition \eqref{condition 1} is satisfied and is not sensitive to the noise or time sampling properties.
In Figure \ref{fig:synthetic1} we inject a sinusoidal signal in the data. In each realization the signal phase $\phi \sim \mathcal{U}(0, 1)$ and frequency $1/P \sim \mathcal{U}(1/T,  N / 2T)$ are drawn from a uniform distribution. The amplitude of the injected signal is chosen for each combination of time sampling and noise separately, such that it significantly perturbs the distribution of the test statistic in Figure \ref{fig:synthetic}.
On the injected sinusoidal data, the null-signal template (teal) gives a significantly lower score to the periodic template (orange) and almost perfectly matches the pure-noise simulations with the periodic template (black). This means that Condition \eqref{condition 2} is also satisfied and as before is not sensitive to the noise or time sampling properties. 
We will further test these conditions in the next section with simulations tailored to the SMBHB search in the PTF data.

\section{SMBHB search in PTF data} \label{sec: quasars}
\subsection{PTF Light Curves}
\label{sec:PTF_LCs}
We extract quasar light curves obtained by PTF \citep{law_palomar_2009} and apply the same cuts as in \cite{charisi_population_2016} to obtain a sample of 35,383 spectroscopically confirmed quasars.

Following \cite{charisi_population_2016}, we model the null hypothesis as a damped random walk (DRW) with additional Gaussian measurement noise and a constant offset to represent the mean magnitude.
A DRW is a correlated Gaussian noise with covariance matrix
\begin{equation} \label{eq: drw kernel}
    \Sigma_{ij}^{(DRW)} = \sigma^2 e^{-\vert t_i - t_j\vert / \tau},
\end{equation}
where $\sigma$ is the strength of the DRW noise and $\tau$ its correlation time. We adopt a log-normal prior for both parameters,
\begin{equation} \label{eq: drw prior}
    \log \sigma \sim \mathcal{N}(\log 0.1, 0.2) \qquad \log \tau \sim \mathcal{N}(\log 120\, \text{days}, 0.9),
\end{equation}
following Figure 9 from \cite{charisi_population_2016} and \cite{macleod_modeling_2010}.
The total noise is a combination of the DRW process and the measurement noise, thus the total covariance matrix is
\begin{equation} \label{eq: kernel}
    \Sigma_{ij} = \Sigma_{ij}^{(DRW)} + \sigma_i^2 \delta_{ij}.
\end{equation}

We first remove the outliers as follows: we identify the data point $i$ as an outlier if it deviates from the Gaussian Process (GP) fit to the data by more than $3$ standard deviations,
\begin{equation}
    \vert x_i - \mu_{GP}(t_i)\vert > 3 \sqrt{\sigma_i^2 + \sigma^2_{GP}(t_i)}.
\end{equation}
We take the GP with the kernel from Eq. \eqref{eq: kernel} with fixed $\sigma = 0.1$ and $\tau = 100$ days, where $\mu_{GP}$ and $\sigma^2_{GP}$ are the GP posterior mean and variance, respectively.
We iteratively fit the GP to the light curves and identify the outliers for each new fit, repeating the procedure 5 times.
We then bin together the observations taken on the same night, as in \cite{charisi_population_2016}.

\subsection{Signal prior} \label{sec: prior}

The alternative hypothesis ($\mathcal{H}_1$) is a sinusoidal signal with period $P$. In the quasar reference frame the period is expected to be distributed as $p(P_{RF}) \propto P_{RF}^{\alpha}$, with $\alpha = 8/3$, assuming that SMBHBs are inspiraling on circular orbits and their evolution is driven by gravitational waves only \citep{haiman_population_2009}.
The observed period is related to the reference frame period by $P = P_{RF} (1+z)$, where $z$ is the redshift of the quasar. Therefore the observed period is also distributed as $p(P) \propto P^{\alpha}$. 
We will enforce an upper limit on the period of $P < P_{\mathrm{cut}} = T/n_{\mathrm{min}}$ to ensure that at least a minimal number of signal cycles $n_{\mathrm{min}} = 2$ are observed, otherwise periodicity detection cannot be claimed.\footnote{Note that previous searches imposed a more relaxed requirement of only 1.5 cycles, which may also contribute to the detection of false positives.}
The probability that a given quasar has a signal with a period in the desired range of $0 < P < P_{\text{cut}}$ is
\begin{equation} \label{eq: prior odds}
    P(\mathcal{H}_1) \propto \int_0^{P_{\mathrm{cut}} / (1+z)} p(P_{RF}) d P_{RF} \propto \frac{P_{\mathrm{cut}}^{\alpha + 1}}{(1+z)^{\alpha+1}},
\end{equation}
where $P_{\mathrm{cut}} / (1+z)$ is the period cutoff translated to the quasar reference frame.
The prior odds $P(\mathcal{H}_1) / P(\mathcal{H}_0)$ penalizes quasars with larger redshift and shorter data span, because the required number of observed cycles limits the allowed observed period which translates to a smaller range of reference-frame periods.
The proportionality constant in Eq. \eqref{eq: prior odds} further depends on quasar's brightness, because brighter quasars with more massive putative binaries chirp faster, and so spend less time at some fixed period \citep{peters_gravitational_1964}. We will ignore this effect here and assume the proportionality constant is the same for all quasars. We will fix this constant to a value that ensures $\sum_i P(\mathcal{H}_1 \vert \mathrm{quasar \, i}) = \sum_i P(\mathcal{H}_0 \vert \mathrm{quasar \, i})$. This is in line with the standard practice in hypothesis testing, i.e. that no hypothesis is preferred a priori. The choice of the constant has no effect on the posterior odds as a test statistic, because it just shifts all scores by a constant.
Finally, to make the prior smooth at the edges, we multiply $p(\log P)$ by the cosine-tapered window which has a value of 1 for $P <  P_{\mathrm{min}}/1.2$ and 0 for $P > P_{\mathrm{min}}$. This makes it easier to maximize the posterior density during the computation of the Bayes Factor and also better aligns with our prior beliefs, which dictate a smooth prior.
For the phase we adopt the uniform prior $\phi \sim \mathcal{U}(0, 1)$. For the amplitude we also adopt the uniform prior, but do not specify its upper bound and ignore the trials factor associated with the amplitude. This only amounts to a constant shift of the test statistic which has no relevance for the Bayes Factor as a test statistic \citep{robnik_statistical_2022}.


\subsection{Test statistics} \label{sec: test stat}

\begin{figure*}
    \centering
    \includegraphics[scale = 0.3]{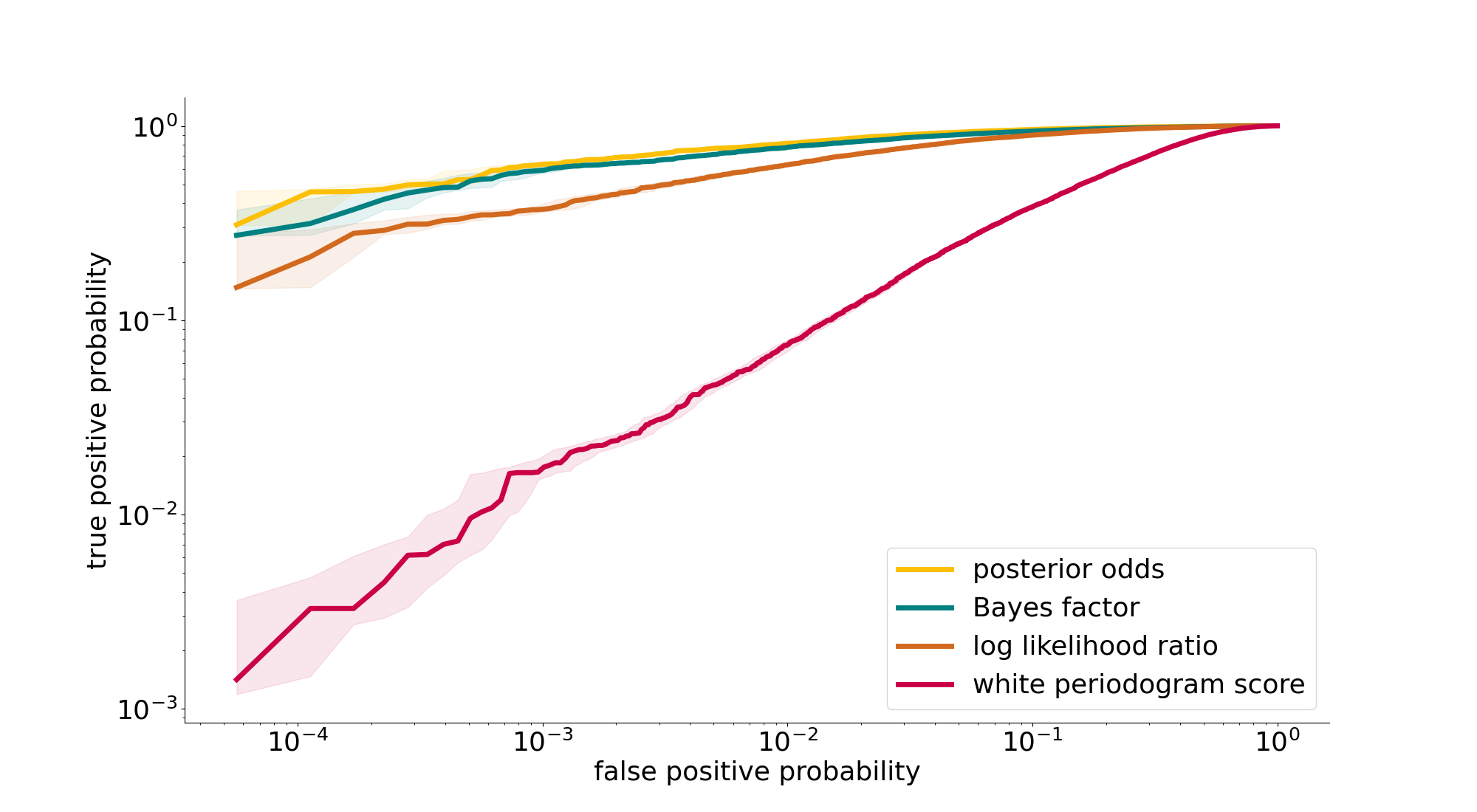}
    \caption{True positive probability as a function of false positive probability (ROC curve) for the periodicity search in the PTF data is shown for the test statistics from Section \ref{sec: test stat}. 
    The shaded regions are $1 \sigma$ confidence bands obtained by bootstrapping.
    The posterior odds and the Bayes Factor perform best and are closely followed by the log-likelihood ratio. White periodogram score performs much worse.
    }
    \label{fig: ROC}
\end{figure*}

We try several test statistics:
\begin{itemize}
    \item Standard (uncorrelated) peridogram score, Eq. \eqref{eq: ls}.
    \item Likelihood ratio, Eq. \eqref{eq: lr}
    \item Bayes Factor, Eq. \eqref{eq: bf}
    \item Posterior odds, Eq. \eqref{eq: po}
\end{itemize}
We show how these can be efficiently computed in Appendices \ref{sec: mf} and \ref{sec: bf}.
In Figure \ref{fig: ROC} we show the receiver operating characteristic (ROC) curve for these test statistics. For this, we generated mock light curves by taking the PTF quasar data, described in Section \ref{sec:PTF_LCs} and replacing the magnitude measurements with realizations of DRW, whose parameters are drawn from the prior, as in Eq. \eqref{eq: drw prior}. A signal with amplitude $A = 2 \sigma$ and phase and period drawn from the priors in Section \ref{sec: prior} is randomly injected in some light curves. The signal is injected according to the prior odds, so in around half of the light curves, but with higher probability where the prior odds are larger. 
To compute the ROC curve for a given test statistic, the detection threshold is varied and at each value the fraction of recovered true signals (i.e. true positive probability) is plotted against the fraction of identified false signals (i.e. false positive probability).

Figure \ref{fig: ROC} shows that the posterior odds (yellow) are optimal in the sense that they maximize the true positive probability at a fixed false positive probability. This is the content of the Neyman-Pearson lemma for the composite hypothesis \citep{zhang_bayesian_2017, fowlie_neymanpearson_2023}, so we have here numerically confirmed it.
Posterior odds are almost indistinguishable from the Bayes Factor (teal), meaning that the prior odds do not play a significant role in this application.
The likelihood ratio (orange), which accounts for the correlated noise, but does not marginalize over the unknown parameters and does not take into account the prior distribution, closely follows the posterior odds and the Bayes Factor but is slightly less optimal. 
The standard periodogram score (red), which assumes white noise, is significantly suboptimal: at 0.001 false positive probability it has a 20 times lower chance of detecting a real signal, because it needs a higher detection threshold to keep the number false positives low. Most of the false positives
arise from the correlated signal of the quasar lightcurves, which this test statistic ignores. 

The take-away message is that the detection efficiency benefits significantly from incorporating the noise correlations in the test statistic.
Because the Bayes Factor and posterior odds perform almost identically and the Bayes Factor is the standard choice, we will use the Bayes Factor in the remainder of the paper. 
Note that one would not benefit significantly in terms of computational cost by using the likelihood-ratio instead. This is because the Gaussian quadrature integration is cheaper than the optimization, which is needed to identify the optimal parameters and is needed for both test statistics.

\subsection{Null-signal template validation}
\begin{figure}
    \centering
    \includegraphics[scale= 0.28]{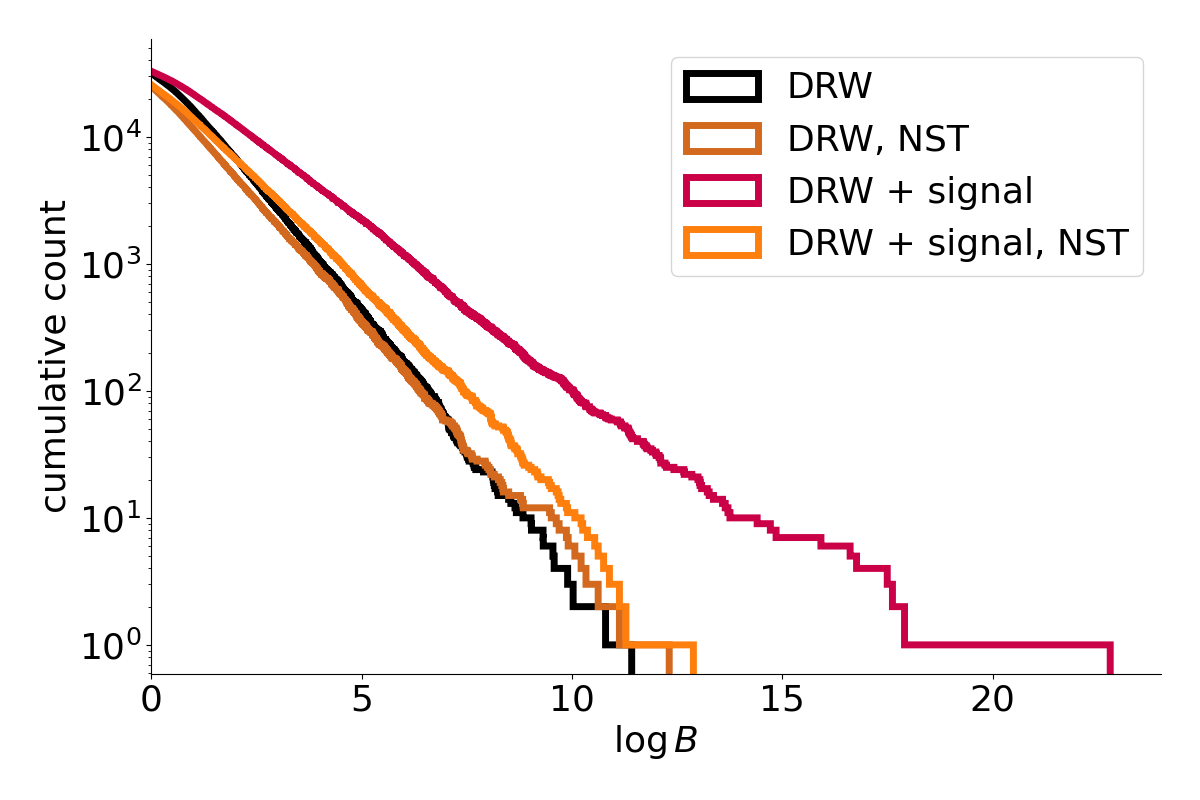}
    \caption{
    Cumulative number of detections as a function of the Bayes Factor test statistic on a simulation of the SMBHB search in PTF data. On noise-only simulation (DRW), the null-signal template (brown) and the periodic template (black) match perfectly, thus Condition \eqref{condition 1} is satisfied. On simulations of the noise with injected signal, the null template (orange) yields significantly lower scores than the periodic template (red) and matches the noise-only simulations. Thus Condition \eqref{condition 2} is also satisfied.
    }
    \label{fig: drw simulations}
\end{figure}

Next we repeat the test for Conditions \eqref{condition 1} and \eqref{condition 2}, as in Section \ref{sec: synthetic}, but specifically for the SMBHB search in PTF. For this, we take the PTF light curves from Section \ref{sec:PTF_LCs} and replace the magnitude measurements with realizations of DRW, whose parameters are drawn from the prior in Eq. \eqref{eq: drw prior}. These light curves serve as simulations of the null hypothesis $\mathcal{H}_0$. As explained in \ref{sec: test stat}, we then use the Bayes Factor as our test statistic.
In Figure \ref{fig: drw simulations} we show that the p-value computations with the periodic signal template (black) and the null-signal template (brown) agree perfectly, thus demonstrating Condition \eqref{condition 1}. We also inject a sinusoidal signal in all light curves, with properties as in the previous section, i.e. with an amplitude of $A = 2 \sigma$, and phase and period drawn from their prior (Section \ref{sec: prior}). This light curves serve as simulations of the alternative hypothesis $\mathcal{H}_1$. Figure \ref{fig: drw simulations} shows that the NST distribution (orange) produces significantly lower scores than the periodic signal (red), and thus Condition \eqref{condition 2} is satisfied.

\subsection{Results}

\begin{figure*}
    \centering
    \includegraphics[scale= 0.35]{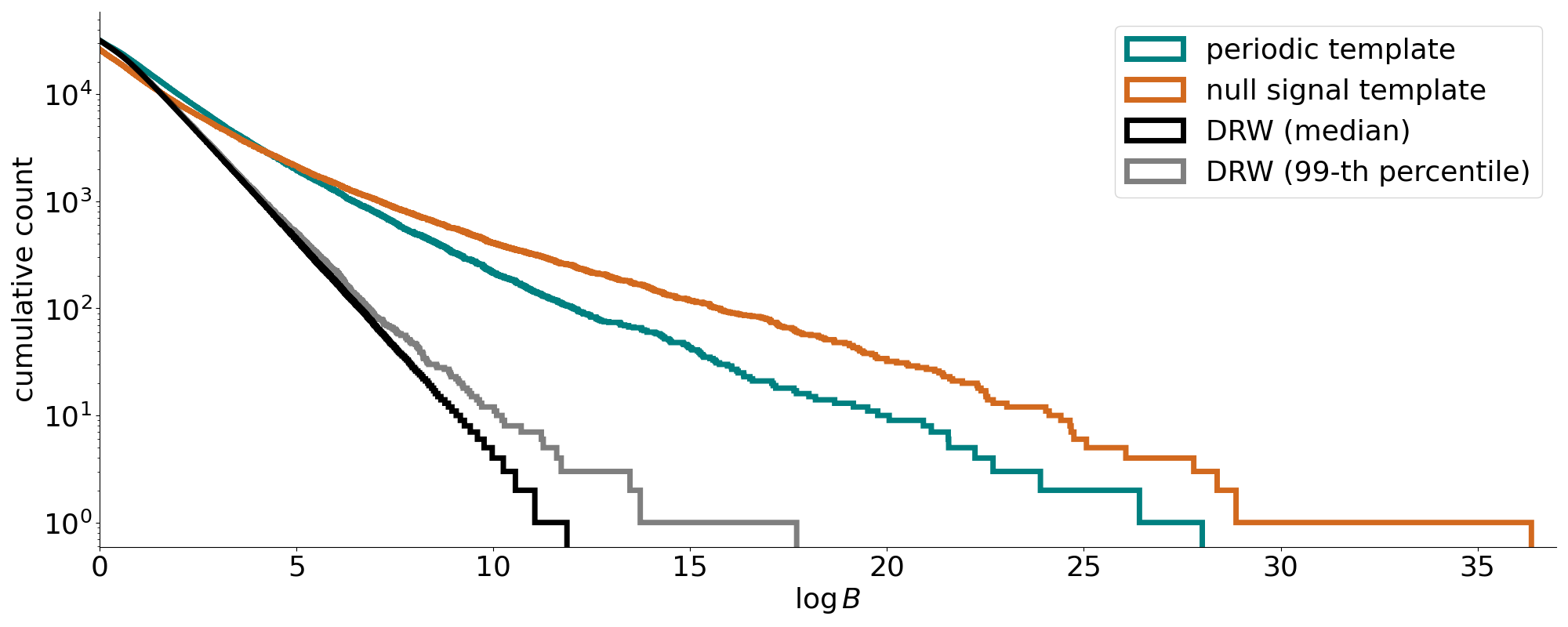}
    \caption{The results of SMBHB search in PTF data. Cumulative count of detections as a function of the detection threshold is shown. The periodic template results (teal) are to be compared with the null-signal template results (orange), which act as a null simulation. Detection does not exceed the null baseline, so a discovery cannot be claimed. In contrast, DRW simulations with parameters drawn from the prior (black) yield significantly lower scores, which would lead to a false signal claim. 
    We have performed 100 simulations of the entire dataset analysis and sorted the candidates in each analysis by their score. The percentiles are then computed for each rank separately.
    }
    \label{fig: results}
\end{figure*}

\begin{figure*}
    \centering
    \includegraphics[scale=0.35]{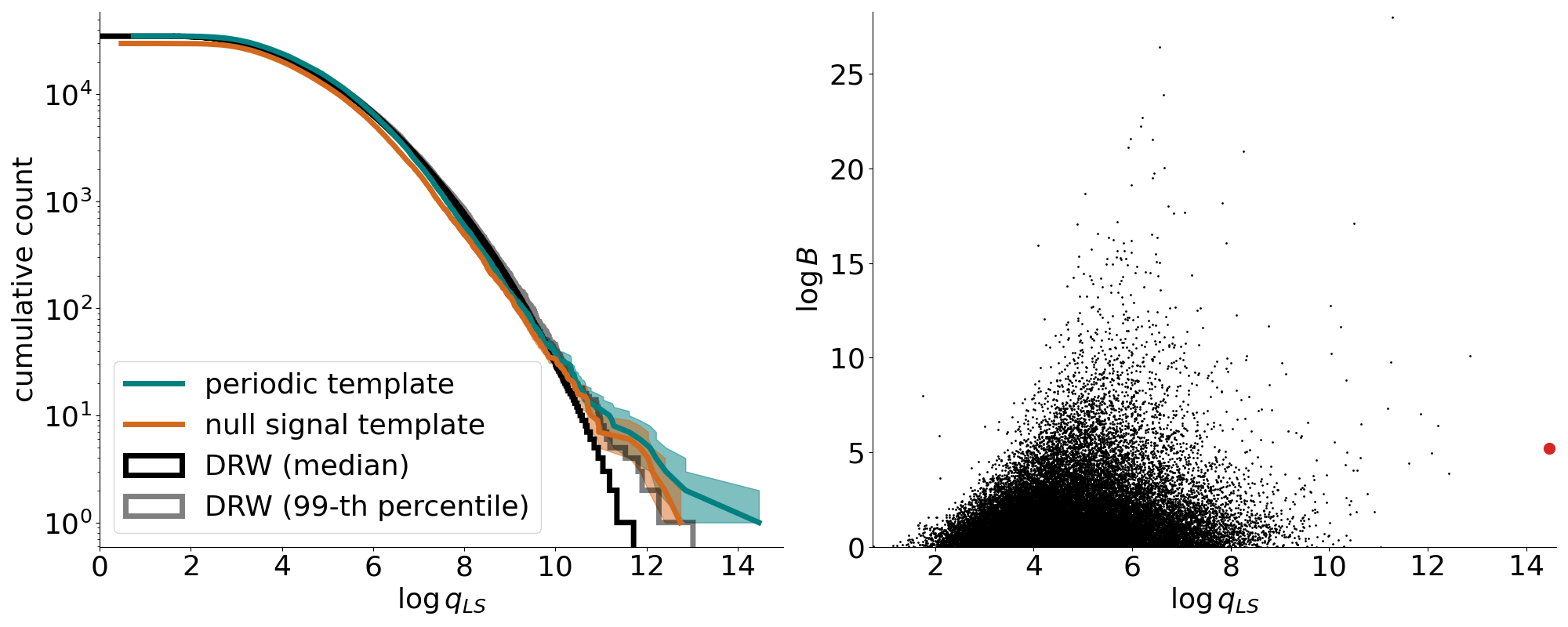}
    \caption{
    Left: The results of SMBHB search in PTF data with the standard white Lomb-Scargle periodogram test statistic $q_{LS}$ from Equation \eqref{eq: ls}. The cumulative count of detections as a function of the detection threshold is shown. The periodic template results (teal) are to be compared with the null-signal template results (orange), which act as a null simulation. Periodic template results do not exceed the null baseline in a statistically significant manner, so a discovery cannot be claimed. The confidence regions are the quartiles obtained by bootstrapping. 
    Here, even the (100) DRW simulations come very close to the null signal template results, showing that the majority of false positives are a result of the test statistic not incorporating the DRW noise.
    Right: the correlation between the white periodogram score and the Bayes factor for all the quasars, using the periodic template. Two test statistics are not highly correlated, especially in the tails: high value of the Bayes factor does not imply high value of the white periodogram score. In particular, the candidate with the highest white periodogram score (marked in red) causes a slight deviation between the NST and periodic template results in the left panel, but has a low value of the Bayes factor. Thus,
    it is most likely caused by DRW not being properly modeled in the white periodogram.
    }
    \label{fig: resultsLS}
\end{figure*}

\begin{figure}
    \centering
    \includegraphics[scale= 0.28]{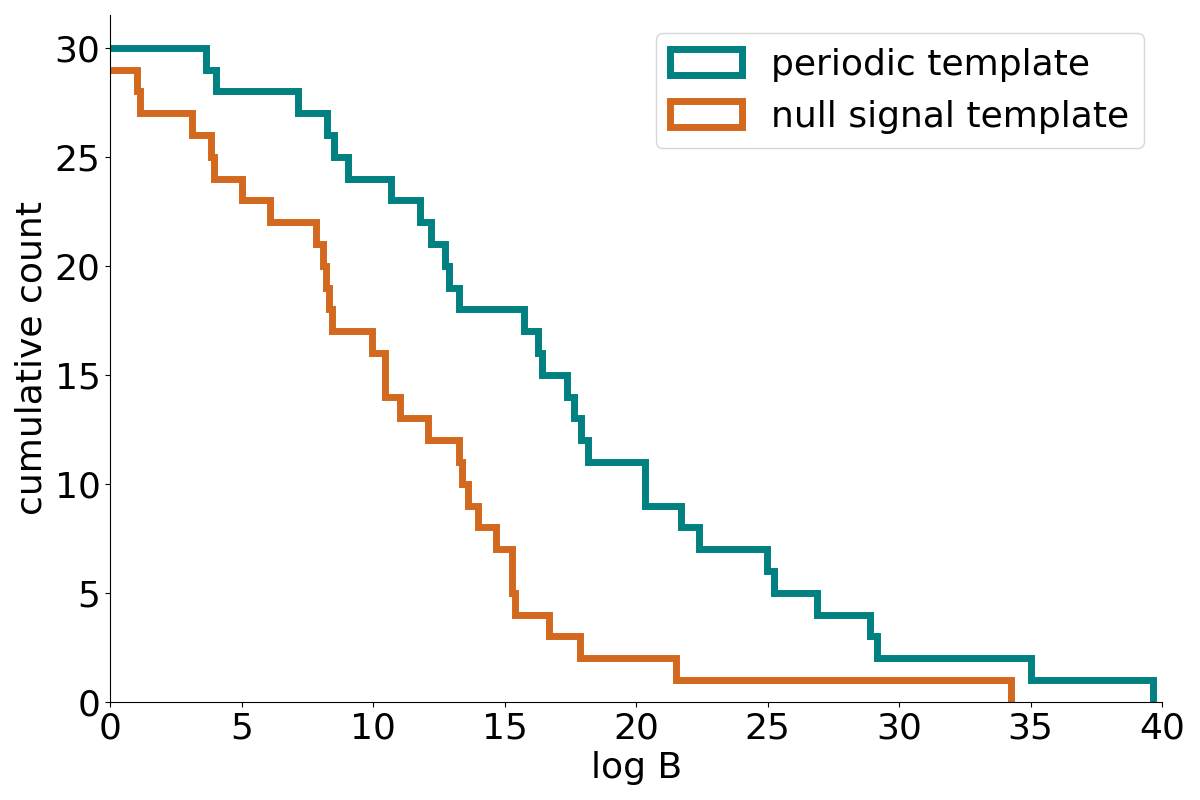}
    \caption{A demonstration that the null-signal template gives a lower significance to the periodic signal. The top 30 quasars from Figure \ref{fig: results} are taken. The reason why these quasars rank so highly might be some corruption of their data, so we extract the optimal signal from these quasars and inject them in 30 among the lower ranking quasars, which will here act as realistic null simulations. To ensure that the signal was injected in a similar environment, we select this second set of 30 quasars to match the original quasars in terms of time span of the data, number of observational nights and parameters of the DRW kernel. In 90\% of cases we were able to match all of those parameters to less than 10\% deviations. 
    We thus obtained 30 simulations of the alternative hypothesis, which is similar to the claimed SMBHB candidates. We analyze these data sets both with the periodic template (teal) and the null-signal template (orange). We show that NST gives lower scores than the periodic template, so such a SMBHB population would have been detectable.
    }
    \label{fig: injected}
\end{figure}

Finally, we apply all the proposed improvements for the periodicity search, in the PTF light curves. In Figures \ref{fig: results} and \ref{fig: resultsLS}, we show the results of the periodicity search, respectively using the Bayes Factor and white periodogram score as the test statistic. All the 33 SMBHB candidates proposed in \cite{charisi_population_2016} achieve $\log B < 11$ so they are no longer among the top candidates when the Bayes Factor is used as a test statistic. The top ranking candidates achieve $\log B > 20$.
This shows that the choice of the test 
statistic has a significant influence 
on the results. We have argued that 
Bayes Factor achieves a significantly 
higher detection sensitivity. 

Based solely on the raw Bayes Factor values, one might be inclined to confirm the top candidates, as even after accounting for the multiplicity of trials, $B \approx \exp{20} / 35 383 \approx 10^4$, where we have treated all quasars equally. This is much larger than $100$, a decisive evidence limit by the Jeffrey's interpretation scale \citep{jeffreys_theory_1998}.
Similarly, the p-values obtained by DRW simulations would suggest a low p-value for these candidates, see Figure \ref{fig: results}. 
However, both of these approaches suffer from poor modeling of the null hypothesis. This is demonstrated with the NST method, which is more robust and agnostic to the assumptions for the underlying noise. Bayes Factors and white periodogram scores, estimated with the NST, reveal that there are no statistically significant periodicity detections, compared to the non-periodic background.  NST reveals a 
much higher Bayes Factors than DRW simulations, demonstrating that the latter are a poor representation of the 
real variability of quasars. 

In fact, NST yields higher Bayes Factor scores, demonstrating that there might be even more non-periodic than periodic signal data. In other words, non-periodic signal would have been a better alternative hypothesis than the periodic signal.
In contrast, we show in Figure \ref{fig: injected} that a real signal would trigger a significantly higher detection with the periodic template than with NST. 
Thus whatever the source of NST  
giving higher Bayes Factor than periodic 
signal is, it cannot be caused by the presence of real 
periodic signal in the data. It is 
however possible that the NST signal 
being above the periodic signal
is caused by the presence of 
quasi periodic signals in the data that 
have templates similar to NST. 
However, investigating this hypothesis would require a more 
detailed analysis which is beyond the 
scope of this paper, where we only 
focus on periodic signals. 

\begin{figure*}
    \centering
    \includegraphics[scale= 0.35]{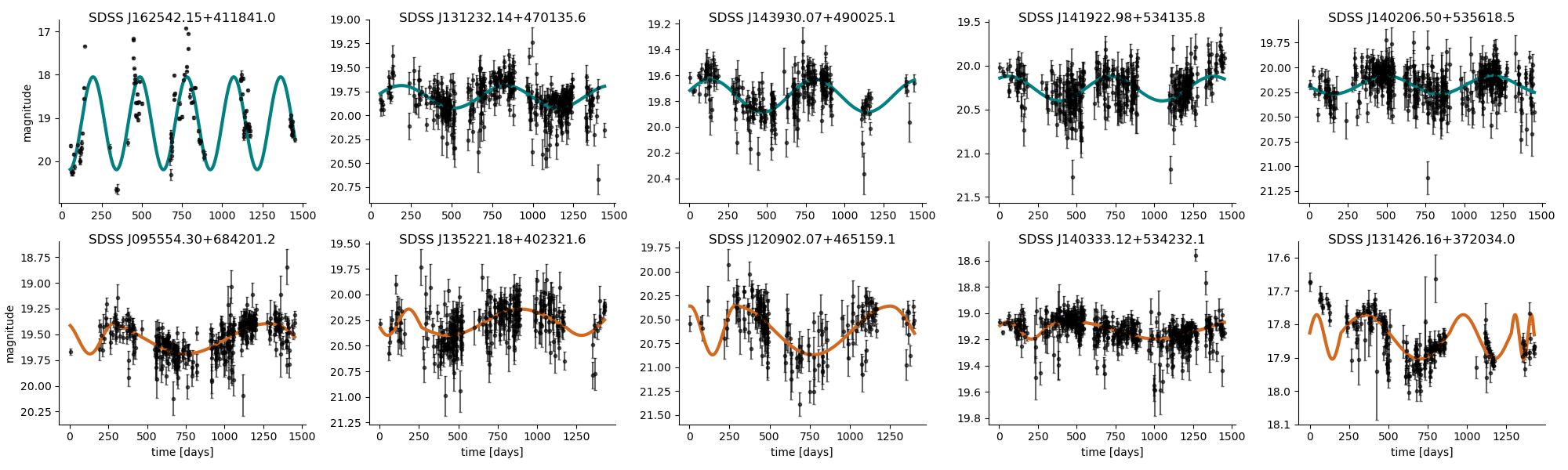}
    \caption{Top: 5 quasars with highest Bayes Factors for the SMBHB hypothesis. Time series data and the best fit models are shown. Time is measured relative to MJD 54903 days. Bottom: same, but with the null-signal template.
    }
    \label{fig: candidates}
\end{figure*}

In Figure \ref{fig: candidates} we show 5 light curves with the highest Bayes Factor, both with the periodic and the null-signal template. In most cases, the
lightcurves are similar, suggesting the
periodogram is triggered on false positives. 


\section{Summary}
We proposed a novel method to improve the accuracy of detecting periodic signals in complex time series data by modifying the periodogram template to create a non-periodic null-signal template (NST). Since anything detected by the NST is by default a false positive, this allows us to use the real data as effective null simulations, avoiding the need for potentially inaccurate data simulations. This is particularly useful in fields like astronomy, where the data complexity and the noise patterns are challenging to model accurately.

We validated our method through synthetic examples and applied it to the search for quasar periodicity, which have been suggested to track the presence of SMBHBs, in $\sim$35,000 quasar light curves from PTF. Our results showed that NST provides a robust estimate of the false positive rate, and revealed that previously proposed SMBHB candidates are likely false positives. The false positive rate in previous analyses was likely underestimated due to the assumption of DRW variability; simulations of DRW light curves cannot reproduce the NST distribution, which indicates that the DRW model may be a limited description of quasar variability, for example that its correlation time may be longer than expected, or that the distribution is non-Gaussian. Since all the systematic searches for quasar periodicity to date assume DRW model for quasar variability \citep{graham_systematic_2015,liu_supermassive_2019,chen_candidate_2020,chen_searching_2024}, many of the identified candidates in these searches may also be false positives, and would need to be validated with the NST method. We plan to apply our method to light curves from other surveys, like CRTS and ZTF.

Additionally, we compared several test statistics (periodogram score, Bayes Factor, posterior odds, and log likelihood ratio). We demonstrated that the Bayes Factor (and posterior odds) outperform the periodogram score, enhancing the sensitivity of periodicity searches by a factor of 20. This indicates that including the correlated noise in the test statistic is beneficial. In the future work the test statistic could have been improved further, by computing the Bayes Factor against the true null hypothesis, which is not captured by DRW, as suggested by our results.

We also introduced a faster algorithm for computing the Bayes Factor of sinusoidal SMBHB signals against the DRW quasar noise, enabling its application to large datasets, like the upcoming LSST of the Rubin Observatory. 

Looking towards future applications of NST, 
it is worth repeating the analysis with 
non-sinusodial periodic templates, as 
well as with quasi-periodic templates. 
The challenge of these applications is 
again to find NSTs that satisfy 
conditions \ref{condition 1} and \ref{condition 2}.
Furthermore, while we have applied NST to a problem concerning SMBHB, the method will be useful in any application where one is searching for periodic signals, for example in exoplanet detection.


\section*{Acknowledgements}
This material is based upon work supported in part by the Heising-Simons Foundation grant 2021-3282 and by the U.S. Department of Energy, Office of Science, Office of Advanced Scientific Computing Research under Contract No. DE-AC02-05CH11231 at Lawrence Berkeley National Laboratory to enable research for Data-intensive Machine Learning and Analysis. MC is funded by the European Union (ERC-STG-2023, MMMonsters, 101117624).  ZH acknowledges support by NSF grant AST-2006176 and NASA grants 80NSSC22K0822 and 80NSSC24K0440.

\section*{Data Availability}
The data underlying this article will be shared on reasonable request to the corresponding author.

\appendix

\section{Periodogram as a matched filter} \label{sec: mf}

The likelihood of the Gaussian data under the null hypothesis is
\begin{equation}
    - 2 \log p (\x \vert \Sigma) = \x^T \Sigma^{-1} \x + \log \det 2 \pi \Sigma
\end{equation}
and
\begin{equation}
    -2 \log p( \x \vert \boldsymbol{z}, \Sigma) = -2 \log p( \x - \boldsymbol{s}(\boldsymbol{z}) \vert \Sigma) 
\end{equation}
under the alternative hypothesis.
$\Sigma$ is the noise covariance matrix which we will assume to be known in this section.
$\boldsymbol{z}$ are the parameters of the signal. Let the signal depend on the first $M$ parameters linearly, we will call these parameters the amplitudes and denote them by $\boldsymbol{a}$.
The signal template is then of the form 
\begin{equation} \label{eq: template}
    \s = \sum_{i = 1}^M a^i \s_i (\boldsymbol{z}_{>M}) \equiv a^i \s_i (\boldsymbol{z}_{>M}).
\end{equation}
As in the second step, Einstein's convention will be used from now on: whenever there are repeated indices, the sum $\sum_{1}^M$ is implied.

We are mainly be interested in the floating mean periodogram, where
\begin{align}
    \boldsymbol{s}_1(t) &= \sin 2 \pi \, t / P  \\ \nonumber
    \boldsymbol{s}_2(t) &= \cos 2 \pi \, t / P \\ \nonumber
    \boldsymbol{s}_3(t) &= 1
\end{align}
and $z_4 = P$ is the period of the signal.

We would like to maximize the log-likelihood ratio between the two hypotheses
\begin{equation}
    \Delta \chi^2 = 2 \log p( \x \vert \boldsymbol{z}, \Sigma) / p( \x \vert \Sigma) = \braket{\x}{\x} - \braket{\x-\boldsymbol{s}}{\x-\boldsymbol{s}},
\end{equation}
where we have introduced a scalar product $\braket{\boldsymbol{x}}{\boldsymbol{y}} = \boldsymbol{x}^T \Sigma^{-1} \boldsymbol{y}$. 
Let's define the metric 
\begin{equation}
    g_{ij} = \braket{\s_i}{\s_j}
\end{equation}
and its inverse $g^{ij}$, such that $g^{ij} g_{jk} = \delta^i_{\, k}$.
We only consider problems with a low number of linear parameters, so inverting the metric is cheap. 
Let's define the components of the data along the signal vectors as 
\begin{equation}
    x_i = \braket{\x}{\s_i}.
\end{equation}
With this notation the score is
\begin{equation}
    \Delta \chi^2 = 2 a^i x_i - a^i a^j g_{ij}.
\end{equation}

Maximization over the linear parameters can be performed analytically:
\begin{equation}
    \frac{\partial \Delta \chi^2}{\partial a^i} \biggr\rvert_{\widehat{a}} = 2 x_i - 2 g_{ij} \widehat{a}^i  = 0,
\end{equation}
which gives the optimal amplitudes
\begin{equation}
    \widehat{a}^i = g^{ij} x_j.
\end{equation}
The score $\Delta \chi^2$ at the optimal parameters is
\begin{equation} \label{eq: score}
    \Delta \chi^2 = 2 g^{ij} x_i x_j - g^{ij} g^{k l} x_j x_k g_{i k} = g^{ij} x_i x_j. 
\end{equation}
This is the matched filter: the data is matched against the templates $\s_i$ and inverse weighted with the noise power, here included in the definition of the scalar product. Most of the searches 
to date ignore the inverse covariance 
matrix weighting, resulting in a 
sub-optimal matched filter analysis. 

The strategy for finding the optimal parameters $\boldsymbol{z}$ is to compute the score from Equation \eqref{eq: score} on a grid of non-linear parameters and identify the highest score.

The main cost of Equation \eqref{eq: score} is in computing the scalar products $\braket{\cdot}{\cdot}$. If noise is stationary with measurements taken on a regular grid, the computation can be simplified significantly, but we do not make these assumptions here.
The scalar product is computed as
$\braket{\boldsymbol{u}}{\boldsymbol{v}} = \boldsymbol{u} \cdot \widetilde{\boldsymbol{v}}$, where $\cdot$ is the standard scalar product and 
$\widetilde{\boldsymbol{v}}$ is the solution of the linear system
$\Sigma \widetilde{\boldsymbol{v}} = \boldsymbol{v}$.
Since the covariance matrix is positive definite we can decompose it using a Cholesky decomposition as $\Sigma= L L^T$, where $L$ is lower triangular. Once the Cholesky decomposition is computed, solving the linear system is easy: first one solves $L \boldsymbol{v}' = \boldsymbol{v}$ for $\boldsymbol{v}'$ and then $L^T \widetilde{\boldsymbol{v}} = \boldsymbol{v}'$ for $\widetilde{\boldsymbol{v}}$.

\section{Efficient Bayes Factor computation} \label{sec: bf}
The difficulty in computing the Bayes Factor (Equation \eqref{eq: bf}) is in doing the evidence integrals 
\begin{equation} \label{eq: evidence}
     p(X \vert \mathcal{H}_i) = \int p(X \vert \boldsymbol{z}_i) p(\boldsymbol{z}) d \boldsymbol{z}_i.
\end{equation}
Linear parameters have a Gaussian posterior so can be integrated out of Equation \eqref{eq: evidence} analytically. They contribute a constant which depends only on the prior volume of the linear parameters and will therefore be irrelevant if the Bayes Factor is not taken at a face value, but as a test statistic \citep{robnik_statistical_2022}. What remains is therefore to integrate over the non-linear parameters at optimal amplitude parameters $\boldsymbol{a}(\x, \boldsymbol{z}_{> M})$:
\begin{align} \label{eq: modified evidence}
    \widetilde{p}(x \vert \mathcal{H}_1) = \int p(\x \vert \widehat{\boldsymbol{a}}(\x, \boldsymbol{z}_{> M}), \boldsymbol{z}_{> M}) p(\boldsymbol{z}_{> M}) d \boldsymbol{z}_{> M}.
\end{align}
Because the number of non-linear parameters is low, we can avoid the cost of MCMC by using the Gaussian quadrature scheme. 
Our strategy is to find the parameters which maximize the posterior density and compute the Hessian of the negative log posterior at those parameters. This is the Lapalace approximation to the posterior. It is used to set out the Gaussian quadrature scheme and integrate the posterior \citep{robnik_statistical_2022}.
In Section \ref{sec: quasars}, the null has two non-linear parameters and the alternative has one additional non-linear parameter. We will use quadrature schemes of order 6 and 7 from \cite{stroud_approximate_1963}. We have compared these computations with pocoMC \citep{karamanis_pocomc_2022, karamanis_accelerating_2022}, which is a state of the art MCMC method. Both methods agree to within the MCMC error but quadrature uses 12 log-posterior evaluations in two dimensions and 27 evaluations in three dimensions, while pocoMC uses $2 \times 10^5$ in its default setting and still several thousand calls if specialized to the simple low dimensional targets.

Thus the procedure for computing the Bayes Factor is:
\begin{enumerate}
    \item Maximize the null log-posterior to find the optimal patamers of the null and the Hessian at the peak.
    \item Using the Gaussian quadrature scheme, compute the evidence for the null.
    \item Take the optimal null parameters in the matched filter periodogram from Section \ref{sec: mf} and find the optimal parameters of the alternative hypothesis.
    \item Maximize the alternative hypothesis log-posterior over all parameters of the alternative hypothesis (including the null parameters).
    \item Using the Gaussian quadrature scheme, compute the evidence for the alternative.
\end{enumerate}

\section{Interpolation from the null to the periodic template}

\begin{figure*}
    \centering
    \includegraphics[scale=0.35]{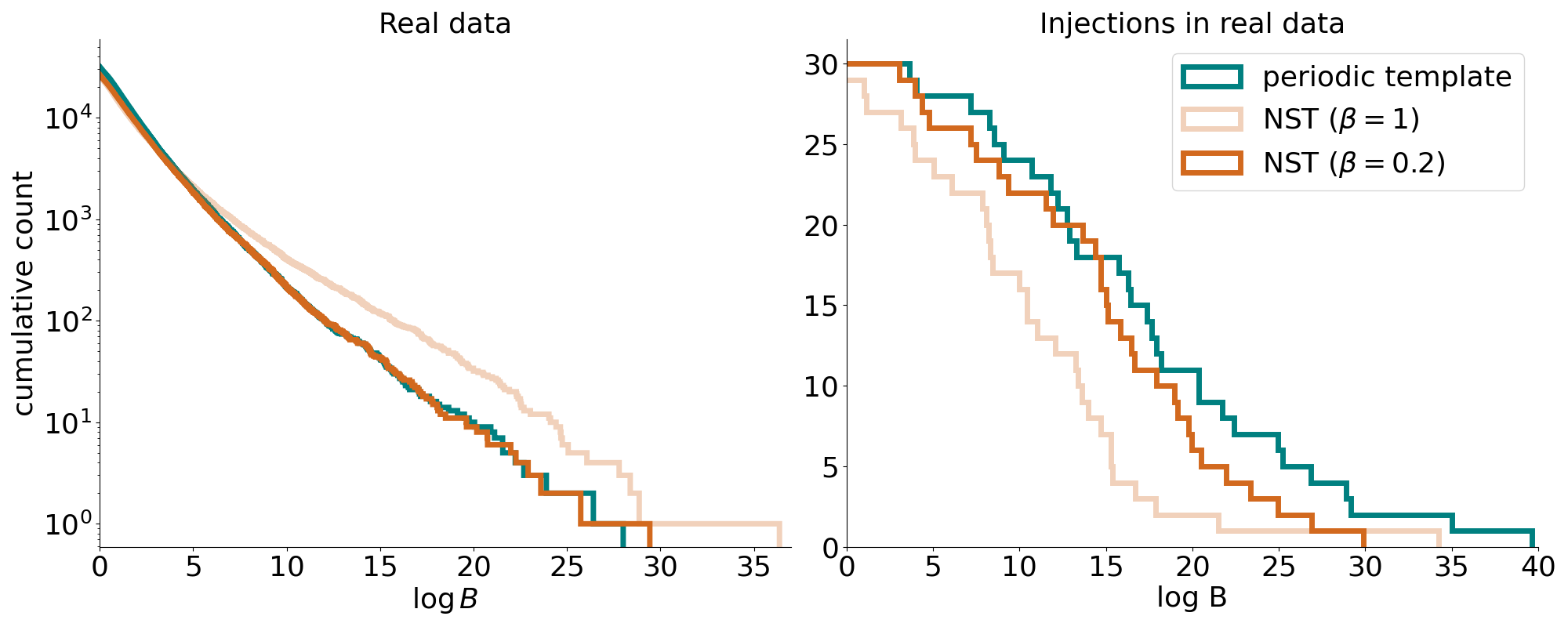}
    \caption{Detection count as a function of the Bayes Factor detection threshold. 
     Left: periodic template and the null-signal template are compared on real data, as in Figure \ref{fig: results}, but with the addition of the $\beta = 0.2$ NST. 
     $\beta = 0.2$ template yields a distribution which is practically indistinguishable from the distribution with the periodic template.
     Right: same but with periodic signal injection in the real data, equivalent to Figure \ref{fig: injected} in the main text. Even at $\beta = 0.2$, NST yields noticably lower scores than the periodic template, suggesting that detections in the left panel are false.
    }
    \label{fig:mixture}
\end{figure*}

In Figure \ref{fig: results} we saw that the NST yields higher scores than the periodic template. This suggests that there is more non-periodic than periodic features in the data. As an additional test we here make the NST perturbation from periodicity less drastic.
We introduce a family of null-signal templates, parametrized by the mixture parameter $\beta$. NSTs in this family are of the form described in Section \ref{sec: method} and have 
\begin{equation}
    p_n = \beta \, p_n^{\mathrm{(optimal)}} + (1-\beta),
\end{equation}
where $p_n^{\mathrm{(optimal)}}$ are the values from Section \ref{sec: method}. Thus this family smoothly interpolates between the periodic template ($\beta = 0$) and the NST from Section \ref{sec: method} ($\beta = 1$). In Figure \ref{fig:mixture} we reanalyze the real data (equivalent of Figure \ref{fig: results} in the main text) and real data with injected signal 
(equivalent of Figure \ref{fig: injected} in the main text) with NST that has $\beta = 0.2$. The real data analysis shows that the NST gives the same test statistic distribution as the periodic template. The injected data demonstrates that if the real signal is present in the data, NST yields lower scores.

\bibliographystyle{mnras}
\bibliography{references}

\bsp	
\label{lastpage}
\end{document}